\documentclass[nofootinbib,twocolumn,aps,pre,superscriptaddress,citeautoscript,floatfix]{revtex4-2}
\usepackage[bookmarks=false]{hyperref}
\usepackage{graphics}
\usepackage{graphicx}
\usepackage{color}
\usepackage{amsmath}
\usepackage{amssymb}
\usepackage{float}
\newcommand{\be}{\begin{equation}}
\newcommand{\ee}{\end{equation}}
\newcommand{\bea}{\begin{eqnarray}}
\newcommand{\eea}{\end{eqnarray}}
\newcommand{\kB}{k_{\rm B}}

\begin{document}
\title{Diffusive dynamics of charge regulated macro-ion solutions}
\author{Bin Zheng}
\email{zhengbin@ucas.ac.cn}
\affiliation{Wenzhou Institute, University of Chinese Academy of Sciences, Wenzhou, Zhejiang 325001, China}
\author{Shigeyuki Komura}
\email{komura@wiucas.ac.cn}
\affiliation{Wenzhou Institute, University of Chinese Academy of Sciences, Wenzhou, Zhejiang 325001, China}
\affiliation{Oujiang Laboratory, Wenzhou, Zhejiang 325000, China}
\affiliation{Department of Chemistry, Graduate School of Science, Tokyo Metropolitan University, Tokyo 192-0397, Japan}
\author{David Andelman}
\email{andelman@post.tau.ac.il}
\affiliation{School of Physics and Astronomy, Tel Aviv University, Ramat Aviv 6997801, Tel Aviv, Israel}
\affiliation{Center of Physics and Chemistry of Living Systems, Tel Aviv University, Ramat Aviv 6997801, Tel Aviv, Israel}
\author{Rudolf Podgornik}
\email{podgornikrudolf@ucas.ac.cn}
\affiliation{School of Physical Sciences and Kavli Institute for Theoretical Sciences, University of Chinese Academy of Sciences, 
Beijing 100049, China}
\affiliation{Wenzhou Institute, University of Chinese Academy of Sciences, Wenzhou, Zhejiang 325001, China}
\affiliation{CAS Key Laboratory of Soft Matter Physics, Institute of Physics, Chinese Academy of Sciences, Beijing 100190, China}

\begin{abstract}
Onsager's variational principle is generalized to address the diffusive dynamics of an electrolyte solution composed of charge-regulated macro-ions and counterions.  
The free energy entering the Rayleighian corresponds to the Poisson-Boltzmann theory augmented by the charge-regulation mechanism. 
The dynamical equations obtained by minimizing the Rayleighian include the classical Poisson-Nernst-Planck equations, the Debye-Falkenhagen 
equation, and their modifications in the presence of charge regulation. By analyzing the steady state, we show that the charge regulation has an important 
impact on the non-equilibrium macro-ion spatial distribution and their effective charge, deviating significantly from their equilibrium values. Our model, based on 
Onsager's variational principle offers a unified approach to the diffusive dynamics of electrolytes containing components that undergo various charge association/dissociation processes. 

\end{abstract}

\maketitle

{\bf Introduction:} Charged macro-ions in solution do not keep their charge fixed but rather respond to the local 
environment by modifying their surface charge density and surface potential,  depending on their local concentration 
and the bathing solution conditions~\cite{Lund_2013, markovich2021charged, avni2019charge}. This conceptual 
framework is called {\it charge regulation} (CR), encompassing charging equilibria of macromolecules in ionic solutions. 
It is ubiquitous and governs important aspects of electrostatic interactions in biological systems~\cite{Zhou2018,Anze2021}. 

The CR phenomenon is essential in understanding how proteins and charged biomolecules change their state via charge 
association/dissociation processes~\cite{ong2020modeling} involving ions in solutions~\cite{borkovec2001ionization}. In particular, 
it affects polyelectrolytes that undergo protonation/deprotonation reactions on acidic/basic sites~\cite{borukhov2000polyelectrolyte,Blossey2023}, 
protein complexation~\cite{da2018protein}, polyelectrolyte gel swelling~\cite{zheng2021phase}, adsorption of charge particles onto surfaces~\cite{hyltegren2017adsorption,Sivan2022}, 
bacterial adhesion~\cite{hong2008electrostatic}, viral capsids assembly~\cite{nap2014role}, zwitterionic colloids and nanoparticles~\cite{Ong2020, Yuan2022}, as well as many other bio-processes. 

Equilibrium CR effects have been extensively studied by including the association/dissociation equilibrium into the mean-field 
Poisson-Boltzmann (PB) theory~\cite{avni2019charge}. However, despite the large progress in the study of equilibrium 
CR phenomena~\cite{avni2019charge},  starting from the seminal work of Ninham and Parsegian~\cite{ninham1971electrostatic}, 
a theoretical understanding of {\it dynamical} CR behavior is less developed. Nevertheless, the latter has pronounced importance 
in numerous physical and chemical processes, such as the kinetics of surfactant adsorption at the air/water interface~\cite{diamant1996kinetics,werkhoven2019dynamic}, 
interactions and dynamics of colloids at the oil/water interface~\cite{Everts2017}, and ionic conductance through nano-tubes~\cite{biesheuvel2016analysis} and nano-channels~\cite{jiang2011charge, Bazant2022}. 

Conventional theoretical studies of charged macro-ion dynamics driven by external electric fields are typically based on the 
Poisson-Nernst-Planck (PNP) theory~\cite{bazant2009}. This theory is a diffusive kinetic extension of the PB formulation of electrostatics. It has been generalized to include ion-ion interactions and steric effects~\cite{kilic2007steric}. 
However, a complete theory of CR dynamics would need even further modifications. It should include a description of the charge dissociation 
processes~\cite{Everts2017, avni2018charge,avni2020critical}, either on the system bounding surfaces~\cite{podgornik2018general} or on the surface 
of the mobile macro-ions~\cite{markovich2018complex} containing the dissociable moieties. In order to formulate these ideas into a consistent theoretical description, 
we chose the framework provided by Onsager's variational principle (OVP)~\cite{doi2011onsager,DOI2021101339}. 
This principle has already proved quintessential when addressing the dynamics of other soft matter systems~\cite{arroyo2018onsager, D0SM02076A, Komura2023}.  

OVP allows us to combine the CR theory already studied in thermodynamic equilibrium~\cite{avni2019charge} with the non-equilibrium dissipation phenomenology, 
as represented by diffusion currents~\cite{xu2023coupled}, charge currents~\cite{gu2018stochastic} and chemical reaction kinetics~\cite{bazant2013theory}. By generalizing OVP even further, 
and including the CR equilibrium free energy and its corresponding diffusive-current densities, we derive the Rayleighian that contains the CR diffusive components. 
Furthermore, our augmented theory yields a set of diffusive dynamic equations. They reduce, in the limit of fixed ionic charge, to the PNP~\cite{gavish2018solvent,Everts2017} 
and Debye-Falkenhagen~\cite{janssen2018transient} equations. We explicitly solve these modified PNP diffusive-dynamic equations  
in the steady-state limit~\cite{Bier_2024} and show that the CR significantly influences the spatial distribution and charge density in externally driven systems. 
There is a clear advantage in formulating the CR dynamics based on OVP. It presents a universal approach for deriving the CR diffusive dynamics directly 
from the equilibrium free energy while making it applicable to various CR models with potential implications for biological systems.

\begin{figure}[h!]
		{\includegraphics[width=0.50\textwidth,draft=false]{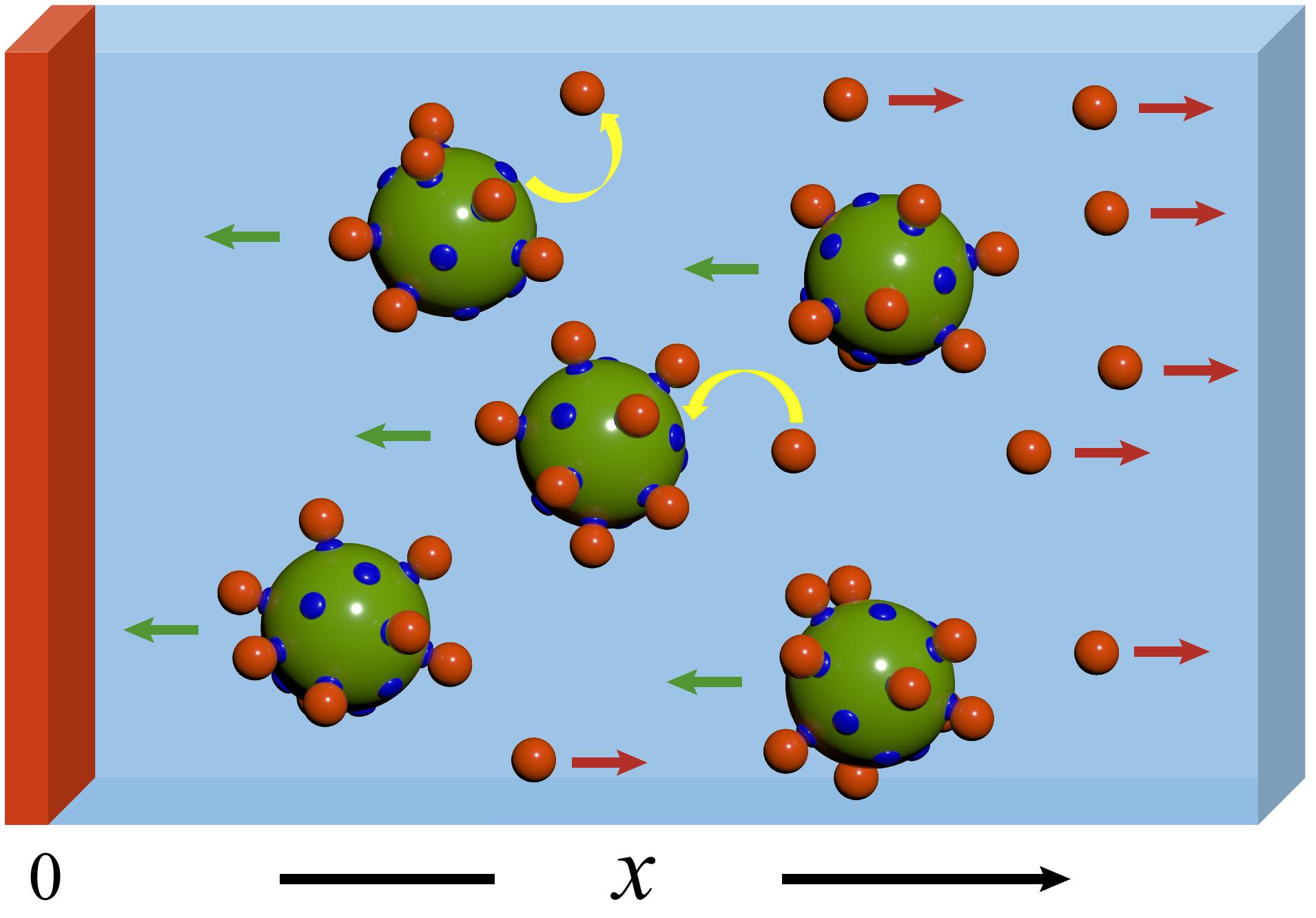}}
	\caption{
		Schematic presentation of our CR diffusive dynamical model. A positively charged wall (red) induces an external electric field, 
		and is placed in contact with a semi-infinite ionic solution. The solution contains negatively charged macro-ions (green) 
		and monovalent positive counter-ions (${\rm B^{+}}$, red) of bulk concentration $n_{\rm b}$ and $p_{\rm b}$, respectively. 
		Each CR macro-ion contains $N$ negatively charged sites (${\rm A^-}$, blue). However, due to the association/dissociation process, 
		the effective macro-ion charge can vary from $-N e$ to zero. The yellow semicircular arrow corresponds to the association/dissociation reaction in Eq.~\ref{eq_r}.
	}
	\label{fig1}
\end{figure}
\hfill \break

{\bf Equilibrium CR model:} We consider a positively charged planar boundary placed at $x=0$. This plane induces 
a static electric field on a semi-infinite ionic solution, as shown in Fig.~\ref{fig1}. The solution is composed of negatively 
charged macro-ions of spatially varying concentration $n({\bf r})$ and bulk value $n_{\rm b}$, and positively charged counter-ions 
concentration $p({\bf r})$ of bulk value  $p_{\rm b}$ (denoted as ${\rm B^+}$). Each macro-ion contains $N$ negatively 
nominally charged sites (denoted as ${\rm A^-}$), and each of the ${\rm A^-}$ sites can change its charge by an association/dissociation process, 
\begin{align}
	{\rm A^-}+{\rm B^{+}}\rightleftharpoons {\rm AB}
	\label{eq_r}
\end{align}
The dynamical number fraction of ${\rm A^-}$ sites that are neutralized by ${\rm B^+}$  is $\phi({\bf r})$ and it varies from zero 
(when the macro-ions are fully charged) to unity (when the macro-ions are completely neutral). 

In our model, the overall electro-neutral solution has no co-ions. This requires that the integrated number of ${\rm A^-}$ 
sites is equal to that of ${\rm B^{+}}$. The electro-neutrality condition in bulk can be expressed as $p_{\rm b}=n_{\rm b}N(1-\phi_{\rm b})$, where $\phi_{\rm b}$ 
is the equilibrated number fraction of neutralized ${\rm A^-}$ sites in the bulk.

Within the mean-field framework, the thermodynamic free energy $F$ is a sum of the electrostatic free energy, the mobile ion translational entropy term $TS(p,n)$, 
and the CR free energy per macro-ion $g(\phi)$. Hence, $F$ can be written as~\cite{markovich2018complex, avni2018charge} 

\be
	\begin{aligned}
		F[\psi, p, n, \phi] =&\int f(\psi, p, n, \phi)\,{\rm d}^3 r \\
		= &\int \Bigg( -\frac{\varepsilon}{2}(\nabla \psi)^2+e \psi\left[p-nN(1-\phi)\right] \\
		&+TS(p,n) + ng(\phi) \Bigg) {\rm d}^3 r , 
	\end{aligned}
\label{Totalenergy}
\ee
where $\psi({\bf r})$ is the electrostatic potential, $T$ is the temperature, $\varepsilon=\varepsilon_0 \varepsilon_r$ is the dielectric constant of the solution, 
$\varepsilon_0$ is the vacuum permittivity, $\varepsilon_r$ is the relative permittivity, and $e$ is the elementary charge. Furthermore, 
\be
	\begin{aligned}
S(p,n) =\kB  \bigg(p\left[\ln \left(p a^3\right)-1\right]+n\left[\ln \left(n a^3\right)-1\right]\bigg) 
	\end{aligned}
\ee
is the mixing entropy of counter-ions and macro-ions in the dilute solution limit, and $\kB $ is the Boltzmann constant. 
For simplicity, the molecular size difference is ignored, and both macro-ions and counter-ions are assumed to have the same molecular volume, $a^3$. 

The CR model is based on the Langmuir isotherm describing the charge association/dissociation process~\cite{podgornik2018general}. 
Within this framework, $\phi$ is an annealed variable and the CR free-energy density $g(\phi)$ is given by
\be
	\begin{aligned}
		g(\phi)=N \Big(\alpha\,\phi+\kB T \left[\phi \ln{\phi}+(1- \phi)\ln\left(1- \phi\right)\right]\Big),
	\end{aligned}
\ee
where $\alpha$ is the association/dissociation parameter, and the last two terms correspond to the mixing entropy of $N$ adsorption sites on each macro-ion. 
We note that other models can be applied to account for generalized CR processes~\cite{avni2020critical,borkovec2001ionization}, 
and our approach can include any of them as well~\cite{podgornik2018general}. 

Minimization of the free energy $F$ with respect to $\psi$ leads to the Poisson equation
\be
\nabla^2 \psi=-\frac{e}{\varepsilon}\left[p-nN (1-\phi)\right],
\label{Poisson}
\ee  
while the minimization with respect to the other variables $p$, $n$ and $\phi$ yields the respective chemical potentials. 
Such thermodynamic equilibrium equations for a variant of the above model have already been investigated in Ref.~\cite{markovich2018complex} 
and will not be presented explicitly here.

\hfill \break

{\bf Rayleighian and diffusive currents:} 
Our system contains negatively charged macro-ions in one of the $i=0,..., N$ charge states, each with a number density $n_i$, and counter-ions of density $p$. 
The velocity of each macro-ion ${\bf v}_i$ with magnitude $v_i=|{\bf v}_i|$ depends on its charge state. 
Therefore, there are $N+1$ possible velocities fields of the macro-ions, and the velocity of the counter-ions is denoted as ${\bf v}_p$. The dissipation function $\Phi$ stems from the friction 
in the diffusive motion, as the mobile ions migrate with their respective velocities through the solvent. It is given as

\be
	\begin{aligned}
		\Phi=&\frac{1}{2}\int \left[ \sum\limits_{i=0}^N n_i \xi_i v_i^2\ + \ p\, \xi_{p} v_{p}^2\right]{\rm d}^3r.
	\end{aligned}
	\label{Eq6}
\ee
where $\{\xi_i\}$ and $\xi_{p}$ are the corresponding $N+2$ {\it friction coefficients}. 

In the spirit of the mean-field equilibrium theory, we proceed to simplify the above $\Phi$ by the following assumptions: 
(i) the macro-ions in any of their charge state are moving with the same average velocity $v_i=v$, and (ii) their friction coefficient is proportional to the number of B$^+$ absorbed ions. 
This can be expressed as a linear dependence,  $\xi_i=N\xi_s + i\,\xi_w,$ $i=0,..., N$, 
where $N\xi_s$ is the friction of the macro-ion having no absorbed counter-ions and $i\,\xi_w$ is the added friction for the $i$-th charge state.  
Note that $\sum_{i=0}^N n_i=n$ as $n$ is the total density of the macro-ions. In addition, on the mean-field level, 
we replace  $\sum_{i=0}^N i\, n_i $ by an average over all the $\{i\}$ charge states $ n \langle\, i\,\rangle =nN\phi=w$, 
where $w=s\phi$ and $s=nN$. Also recall that $\phi$ is the number fraction of neutralized sites on the macro-ion. Then, Eq.~(\ref{Eq6}) can be 
simplified and becomes

\be
		\Phi=\frac{1}{2}\int\left( s \xi_{s} v^2 +w \xi_w v^2+ p\, \xi_p v_{p}^2\right){\rm d}^3r.
\ee
	
On the mean-field level, the above equation implies that, it is equivalent to consider that the dissipation comes from three types of mobile components:  
macro-ions that have no B$^{+}$ association with site density $s=Nn$, macro-ions with an average of $N\phi$  
associated B$^+$ counter-ions and density $w=Nn\phi=s\phi$, and free positive counter-ions of density $p$. The velocities of the first two mobile components 
can in principle be defined as $v_s$ and $v_w$, respectively, but we assume that both are equal to $v$.

It is more convenient to express the free energy  $F$,  Eq.~(\ref{Totalenergy}), as $F[\psi, s, w, p]$. 
We now write down the Rayleighian for the three mobile components and employ the Onsager's variational principle (OVP) to derive the dynamical equations~\cite{doi2011onsager, DOI2021101339}. 
The dissipation function can now be rewritten in terms of the respective {\it particle current densities} for each mobile component.  
In terms of the currents defined by  $j_{s} = s v_{s}$, $j_{w} = w v_{w}$ and $j_{p} = p v_{p}$, we have
\be
	\begin{aligned}
		\Phi=&\frac{1}{2\zeta} \int \left[\frac{j_{s}^2}{s}
		+ \frac{j_w^2}{w} + \frac{j_{p}^2}{p}\right]{\rm d}^3r,
	\end{aligned}
\ee
where $\zeta$ is the mobility coefficient, and all three friction coefficients are assumed to be equal,
$\xi_s{=}\xi_w{=}\xi_{p}{=}1/\zeta$.  
Finally, the Rayleighian,  $R=\Phi+\partial_t F$, is composed of the dissipation function $\Phi$ plus the temporal free energy rate 
$\partial_t F=\partial F/\partial t$, and $R$ is written as

\be
	\begin{aligned}
		R=&\,\Phi+\partial_t F \\
		=&\,\Phi+\int \left[\frac{\partial f}{\partial\psi}\frac{\partial\psi}{\partial t}
		+\frac{\partial f}{\partial s}\frac{\partial s}{\partial t}
		+\frac{\partial f}{\partial w}\frac{\partial w}{\partial t}
		+ \frac{\partial f}{\partial p}\frac{\partial p}{\partial t}\right]{\rm d}^3r.
	\end{aligned}
\label{Ray}
\ee
We assume that the electrostatic potential $\psi$ is a ``fast dynamical variable" and that the Poisson equation, Eq.~(\ref{Poisson}), remains valid also in
a close-of-equilibrium situation, $\delta F/\delta \psi=0$. We further assume that the continuity relations always hold for the density variables $s$,  
$w$, and $p$. They connect the time derivative with the divergence of the respective current density, 
\bea
\partial_t k = - \nabla\cdot {\bold j}_k  \qquad {\rm for} \qquad k= s, w, p. 
\label{cont_eq}
\eea
Thus, the terms in the volume integral of the Rayleighian $R$ in Eq.~(\ref{Ray}) can be transformed into purely spatial derivatives. 
The variation of $R$ with respect to the current density variables, $\delta R/\delta j_k=0$ then yields,

\bea
	\begin{aligned}
		{\bold j}_{s}&=-\zeta\left[-es \nabla\psi+\kB T\nabla n - \frac{\kB T s}{1-w/s}\nabla\left(\frac{w}{s}\right) \right] ,\\ 
		{\bold j}_w&=-\zeta\bigg[ew\nabla\psi + \frac{\kB T s}{1-w/s}\nabla\left(\frac{w}{s}\right) \bigg] , \\
		{\bold j}_{p}&=-\zeta\bigg[ ep\nabla\psi+\kB T \nabla p\bigg] .
	\end{aligned}
\label{js}
\eea

A few special cases are of interest.
In the thermodynamic equilibrium, the time derivatives vanish, and we recover 
the equilibrium distribution of ions as was analyzed in Ref.~\cite{markovich2018complex}. 
In addition, Eq.~(\ref{js}) can describe also a steady-state situation, 
which differs from the equilibrium one as it allows for a non-vanishing, 
spatially uniform charge current density \cite{Bier_2024}, as is discussed below.

Furthermore, in the limit of $\phi=0$ and $N=1$ (meaning $s=n$), 
the system contains only monovalent cations and anions.  Equation~(\ref{js}) then reduces to the standard PNP equations 
\be
\begin{aligned}
j_{n}=&-\zeta(-en \nabla\psi+\kB T\nabla n), \\
j_{p}=&-\zeta(ep\nabla\psi+\kB T \nabla p). 
\label{Eq12}
\end{aligned}
\ee
In addition, for the fixed charge (non-CR) case, the charge density is $q=e(p-n)$, and the number density is $\rho=p+n$. Then, 
Eqs.~(\ref{cont_eq}) and (\ref{Eq12}) simplify to 
\be
			\partial_t q =\zeta\left[k_{\rm B} T ~\nabla^2 q + \nabla\cdot (e^2\rho \nabla \psi) \right].
			\label{eq_qt}
\ee
We compute the product divergence in the second term of Eq.~(\ref{eq_qt}) and use the Poisson equation (\ref{Poisson}) for $\nabla^2\psi$. 
To the lowest order in the electrostatic potential with 
$\lambda_{\rm D}^2 = {k_{\rm B}T\varepsilon}/[{e^2}(p + n)] = {k_{\rm B}T\varepsilon}/({e^2}\rho)$,
the above equation becomes $\partial_t q= \kB T \zeta(\nabla^2 q-\lambda_{\rm D}^{-2}q)$, which is exactly the 
Debye-Falkenhagen equation~\cite{janssen2018transient}, describing the dynamics of the charge density.

\begin{figure*}[t]
	{\includegraphics[width=0.95\textwidth,draft=false]{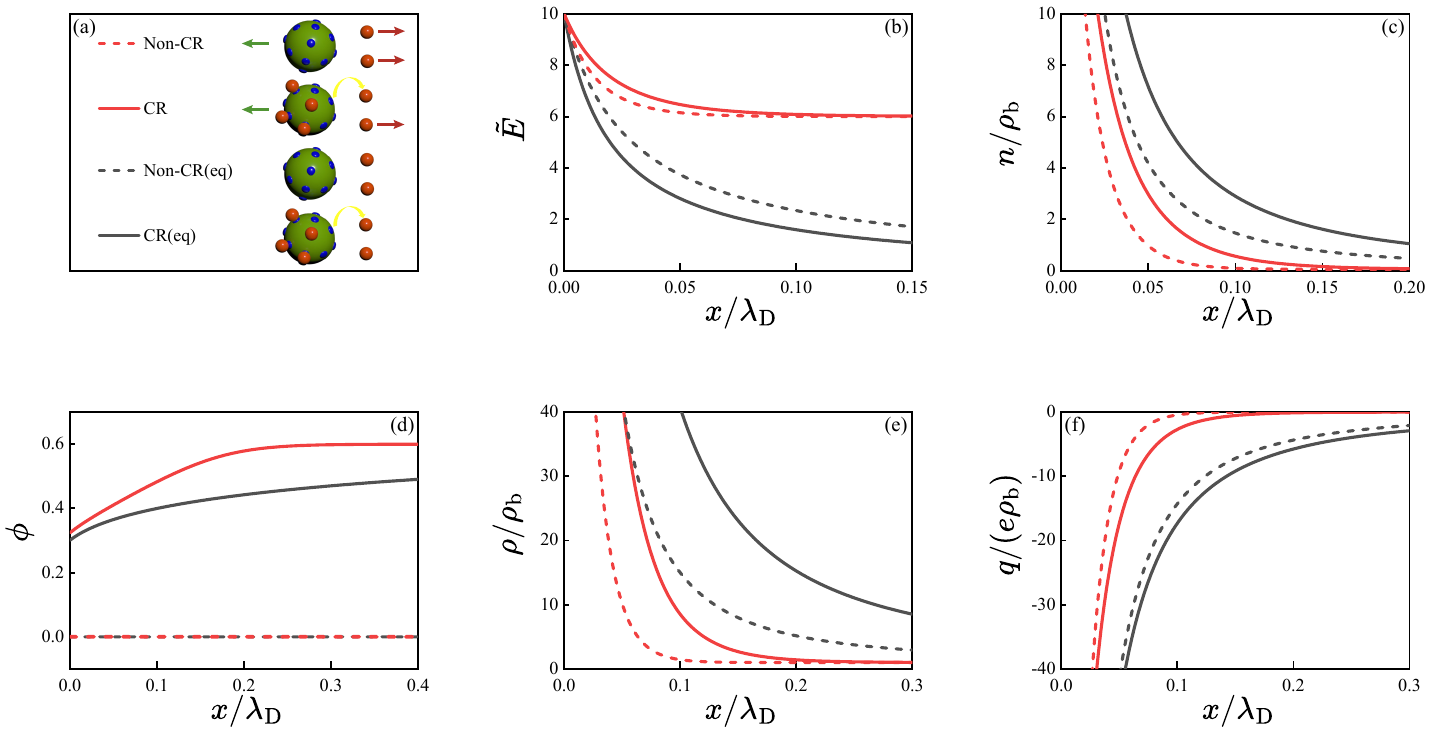}}
	\caption{
		(a) Schematic presentations of the four charge cases. (b) The dimensionless electric field $\tilde E$ (in units of $e \lambda_{\rm D} /\kB T$),  (c) the dimensionless macro-ion density $n/\rho_b$, 
		and (d) the fraction $\phi$. (e) The dimensionless density of the ${\rm A}^-$ sites combined with the total associated and dissociated ${\rm B}^+$ particles $\rho$, 
		and (f) the dimensionless charge density $q/\rho_b$ as a function of $x/\lambda_{\rm D}$, for different values of 
			$\tilde j_q=j_q^{\rm b}\lambda_{\rm D}/(e \kB T \zeta\rho_{\rm b})=0$ ($p_{\rm b}/\rho_{\rm b}{=}0.2$, CR equilibrium case, black line), $\tilde j_q=0$ ($p_{\rm b}/\rho_{\rm b}{=}0.5$, 
			non-CR equilibrium case, black dashed line), $\tilde j_q=6$ ($p_{\rm b}/\rho_{\rm b}{=}0.2$, CR case, red line), and $\tilde j_q=6$ ($p_{\rm b}/\rho_{\rm b}{=}0.5$, non-CR case, red dashed line). 
			The other parameters are  $p_{\rm b}/\rho_{\rm b}{=}0.2$, $N{=}10$,
		$\rho_{\rm b}=2\times10^{-7}{\rm M}$, and $\sigma/\sigma_{\rm sat}{=}2.5$, where $\sigma_{\rm sat}=4\kB T\varepsilon/( e\lambda_{\rm D})$ is the saturation charge density on the wall.
	}
	\label{fig2}
\end{figure*}
\hfill \break
{\bf Steady state:} 
Returning to the CR case, we define
the density of the A$^-$ sites combined with the total associated and dissociated B$^+$ particles as $\rho = s+w+p=Nn(1+\phi)+p$, and the net charge density as $q=e[p-s(1-\phi)]$.  Note that $\rho$ {\it should not} 
be confused with the local number density, $p+n$, and only in the fixed single charge (non-CR and $N=1$) case, $\rho=p+n$ as discussed above.
Additionally, we define the $\rho$ and $q$ conjugate currents: 
${\bold j}_\rho={\bold j}_s + {\bold j}_w + {\bold j}_p$ 
and ${\bold j}_q = e(-{\bold j}_s + {\bold j}_w + {\bold j}_p )$. 

We examine the CR effect in the steady state by setting the time derivatives in Eqs.~(\ref{cont_eq})-(\ref{js}) 
to zero and assuming spatial dependence only in the direction parallel to the external field ($x$-axis). 
This effectively reduces the problem to a one-dimensional one. To maintain a steady state, we assume 
that the total flux of the {\it number} density vanishes $j_\rho=0$, while the net {\it charge} fluxes, $j_q=j_q^{\rm b}$ and $j_w=j_w^{\rm b}$, are constant. 
Hereafter, we use the electric field $E(x)=-\partial_x\psi$ instead of $\psi$, and  
the two ordinary differential equations for  $E(x)$ and $\phi(x)$ can be derived (more details are provided in the S.I.).

The boundary conditions are chosen similarly as by Bier~\cite{Bier_2024}. In the bulk, we stipulate that the electric field $E(\infty)\,{=}\,E_{\rm b}$, 
the number density $\rho(\infty)\,{=}\,\rho_{\rm b}$, $p(\infty)\,{=}\,p_{\rm b}$ and 
from charge neutrality, $\phi(\infty)\,{=}\,\phi_{\rm b}=1-2 p_{\rm b}/\rho_{\rm b}$.
For the boundary condition at $x{=}0$, we choose 
$e E(0)\lambda_{\rm D}/\kB T= 4 \sigma/\sigma_{\rm sat}$, where $\sigma$ is the surface charge density and 
$\sigma_{\rm sat}=4\varepsilon \kB T /(e \lambda_{\rm D})$ is the saturation charge density as defined in Ref.~\cite{fedorov2008towards}. 
Note that a related steady-state case without CR effect was recently analyzed analytically in Ref.~\cite{Bier_2024}; however, the CR model can only be analyzed numerically. 

Thermodynamic equilibrium is characterized by $j_q^{\rm b}=0$, as shown by the solid and dashed black lines 
in Fig.~\ref{fig2}. For non-zero but constant $j_q^{\rm b}$, the system deviates from equilibrium 
into a steady state (the solid and dashed red lines). 
In addition, the CR process can be controlled through the bulk value $\phi_{\rm b}$, with the charge neutrality relation $\phi_{\rm b}=1-2p_{\rm b}/\rho_{\rm b}$.  
Note that $\phi_{\rm b}=0$ or $p_{\rm b}= \rho_{\rm b}/2$ corresponds to a constant maximum charge density on the macro-ion surface. Equivalently, it corresponds 
to the fixed charged (non-CR) case (dashed red and black lines).
Therefore, we present four cases with the equilibrium/steady state and CR/non-CR state combinations in Fig.~\ref{fig2}.
For easier understanding, these four schematic presentations are shown in Fig.~\ref{fig2}(a), respectively.
Figure~\ref{fig2}(b) demonstrates that the electric field $E=-\partial_x \psi$ for the steady state decreases from its surface value to its bulk value for large $x/\lambda_{\rm D}$. 
Hence, the CR process displays small differences compared to the non-CR case (solid vs.\ dashed red line in Fig.~\ref{fig2}(b)).

However, a significant CR effect in the steady state is seen for both the macro-ion concentration profile $n(x)$ and the dimensionless density of the ${\rm A}^-$ sites combined with the total associated 
and dissociated ${\rm B}^+$ particles $\rho(x)$, as shown in Fig.~\ref{fig2}(c) and (e). More negatively charged macro-ions migrate towards the wall due to the electrostatic attraction, as shown in 
Fig.~\ref{fig2}(c). The CR curve (red solid line) shifts significantly to the right, towards larger distances from the wall.  
Thus, the macro-ion density at the same distance from the wall is smaller in the CR steady state than in the equilibrium cases (solid/dashed black lines) but is larger than in the non-CR case.
Additionally, as the macro-ions migrate closer to the wall, more counter-ions dissociate from their surfaces, decreasing $\phi(x)$, as shown in Fig.~\ref{fig2}(d). 
This difference amounts to almost 50 \% in the CR steady state. 

For the non-CR case, we recall that the macro-ions trivially keep a constant charge, i.e., $\phi=0$ (dashed red line in the Fig.~\ref{fig2}). 
The $\rho$ and $q$ plots in Fig.~\ref{fig2}(e) and (f) follow similar tendencies as in $n$ when comparing the four cases. In the counter-ion-only case, 
the distribution of the charge and particle densities are dominated by spatial dependence of macro-ions.


In the steady state, the current density of each component, denoted as $j_{k}, (k=s,w,p)$, has a linear dependency on the bulk value $j_q^{\rm b}$ as charge neutrality 
is obeyed. For example, for the CR case, $j_{p}=(p_{\rm b} /e\rho_{\rm b})j_q^{\rm b}$, 
and this linear dependence slope is different from that of the non-CR one, $j_{p}=j_q^{\rm b}/(2e)$.

While the electric current densities $j_q$, $j_\rho$, and $j_p$ in the steady state should clearly be constant, it is interesting to note that each of their components exhibits a pronounced spatial dependence. 
We now decompose the charge and particle number currents,  $j_{q}$ and $j_{\rho}$, into separate components: $j_{q1}$ and $j_{\rho1}$ are the electric components proportional to the electrostatic field $E=-\partial_x\psi$, and $(j_{q2}, j_{q3}, j_{q4})$ and  $(j_{\rho2}, j_{\rho3}, j_{\rho4})$ are the three diffusive components proportional 
to the respective concentration gradients $\partial_x p, \partial_x \rho$ and $\partial_x q$. The macro-ion current can also be decomposed into only two components, $j_{p1}, j_{p2}$, proportional 
to the electric field and the macro-ion concentration gradient (see Eq. (13) of the S.I.). 

\begin{figure*}[t]
	{\includegraphics[width=0.95\textwidth,draft=false]{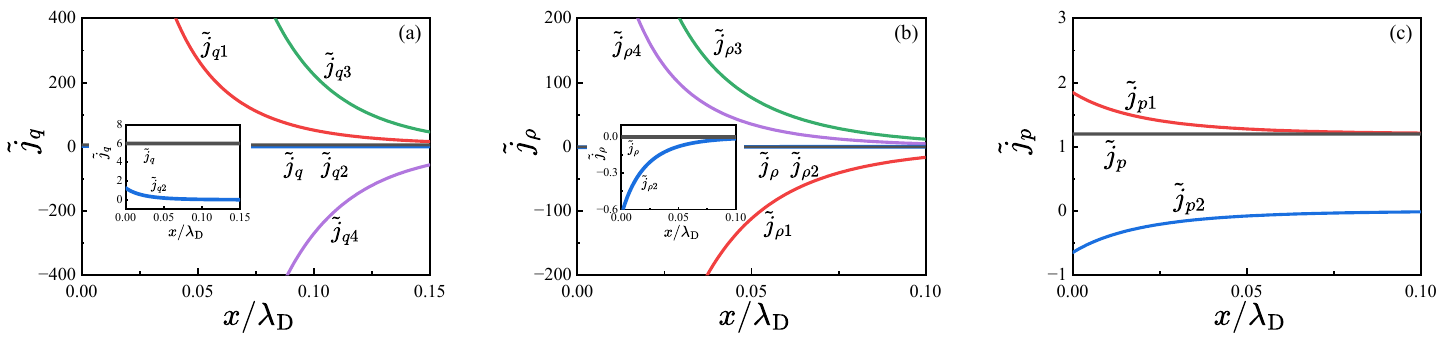}}
	\caption{
		(a) $\tilde j_q$ and its four contributions, (b) $\tilde j_\rho$ and its four contributions, and (c) $\tilde j_p$ and its two contributions, where $\tilde j_q$ is the current $j_q$ rescaled by 
		$\lambda_{\rm D}/(e \kB T \zeta\rho_{\rm b})$, whereas $j_\rho$ and $j_p$ are rescaled by $\lambda_{\rm D}/(\kB T \zeta\rho_{\rm b})$ and denoted as $\tilde j_\rho$ and $\tilde j_p$.
		Other parameters are  $p_{\rm b}/\rho_{\rm b}=0.2$, $N=10$, $\rho_{\rm b}=2\times10^{-7}{\rm M}$, $\sigma/\sigma_{\rm sat}=2.5$, and  $\tilde j_q^{\rm b}=2$. The four contributions 
		in (a) and (b) denoted as $1,...,4$ are the electric component and the three diffusive components, proportional to electrostatic field $-\partial_x \psi$ and concentration gradients $\partial_x p, \partial_x\rho$ and $\partial_x q$ 
		 respectively. Two components in (a) are proportional to $-\partial_x\psi$ and $\partial_x p$ (see Eq.~($13$)-($15$) of the S.I.).
	}
	\label{fig3}
\end{figure*}

The separate spatial dependence of these components is shown in Fig.~\ref{fig3}(a), (b) and (c). Clearly, 
each of the components, $(j_{q2}, j_{q3}, j_{q4})$ and  $(j_{\rho2}, j_{\rho3}, j_{\rho4})$, varies significantly as a function of the distance from the wall, despite their sum remaining constant. 
Additionally, the diffusive components $j_{\rho3}$ and $j_{\rho4}$, corresponding to the density of the A$^-$ sites combined with the total associated and dissociated B$^+$ particles and the net charge density are significantly closer in magnitude than the $j_{q3}$ and $j_{q4}$.

\hfill \break

We have generalized Onsager's variational principle to describe the diffusive dynamics of an ionic solution containing charge-regulated (CR) macro-ions. 
The derived equations represent a consistent generalization of the standard PNP theory that describes fixed charge particles. 
By examining the steady state, we find significant CR effects on the spatial distribution of the macro-ions, particularly in the vicinity of the surface.
Moreover, the electric and diffusive contributions to the current and electric charge densities have pronounced spatial variation, including a significant contribution from the CR components.  

At a fixed distance from the charged surface, the macro-ion density decreases when compared with the equilibrium CR case 
but increases when compared to the steady-state non-CR (fixed charge) case. The CR effects, therefore, always increase the 
macro-ion concentration close to the boundary. In addition, the change in the number of dissociated ions from the macro-ion surface 
is significantly larger in the steady state compared to the equilibrium one, implying that the CR effect becomes stronger in the steady state.

This study employs three assumptions. (i) We assume that the Poisson equation also holds for the slow dynamics considered here, implying that the electrostatic potential is a fast dynamical variable and is always equilibrated.
(ii) The CR process is coupled only to the ionic diffusive dynamics. (iii) The surface charge density 
is kept constant. Our theory provides a unified and consistent way to deal with CR diffusive dynamics for systems undergoing 
charge association/dissociation processes with the bathing solution. Our results and the generalization of Onsager's variational 
principle can provide insight into understanding the transport properties of biomolecular systems, such as proteins and other components of living matter, in the presence of the significant CR mechanism.
\hfill \break
{\bf Acknowledgment:} We thank M. Doi for helpful comments and discussions. BZ acknowledges the National Natural Science Foundation 
of China (NSFC) through Grants No. 22203022, and the Scientific Research Starting Foundation of Wenzhou Institute, UCAS (No.~WIU-CASQD2022016). 
SK acknowledges the NSFC through Grants Nos.~12274098 and 12250710127 and the startup fund of Wenzhou Institute, UCAS (No.~WIU-CASQD2021041). 
DA acknowledges the NSFC-ISF Research Program, jointly funded by the NSFC under grant No. 21961142020 and the Israel Science Foundation (ISF) 
under grants No. 3396/19, and ISF grant No. 226/24. RP acknowledges funding from the NSFC Key Project No.~12034019.

{\bf Supporting Information:}  Detailed derivation for equations in the steady state.

\end{document}